\documentclass{article}
\usepackage{spconf,amsmath,graphicx,hyperref}
\usepackage{url}
\usepackage{bbding} 
\usepackage{multirow}
\usepackage{algorithm}
\usepackage{algorithmic}
\usepackage{colortbl}
\usepackage{cite,xcolor}
\usepackage{diagbox}
\usepackage{booktabs}
\usepackage{amsfonts}
\usepackage{hhline}
\usepackage{subfig}

\usepackage{pifont}

\title{A Unified Uncertainty-Aware Back-End for Speaker Verification: Scoring, Normalization, and Calibration}
\name{Junjie Li$^1$, Kong Aik Lee$^{1,\dagger}$ \thanks{$\dagger$: Corresponding author. }}
\address{$^1$ Department of Electrical and Electronic Engineering,\\ The Hong Kong Polytechnic University, Hong Kong SAR \\
}

\begin{document}
\ninept
\setlength{\textfloatsep}{10pt plus 2pt minus 2pt}
\setlength{\dbltextfloatsep}{10pt plus 2pt minus 2pt}
\maketitle
\begin{abstract}
Speaker verification back-ends commonly combine similarity scoring, score normalization, and calibration. However, speaker embeddings extracted from real-world utterances have trial-dependent reliability because of factors such as duration, noise, and channel variation. Existing uncertainty-aware methods primarily improve the speaker encoder or the initial similarity score, while the estimated uncertainty is typically not propagated through subsequent normalization and calibration. We represent each utterance by a speaker embedding, interpreted as a posterior mean, together with its covariance as an uncertainty estimate. We present a unified uncertainty-aware back-end comprising uncertainty-aware cosine scoring, uncertainty-aware AS-Norm (UAS-Norm), and uncertainty-aware Quality Measure Function calibration (UQMF). Covariance information is incorporated throughout this pipeline to adjust score scaling, cohort statistics, normalized-score combination, and calibration features. Experiments with ECAPA-TDNN and ResNet show consistent EER reductions and improved target--non-target separation across both architectures.\footnote{The official implementation is available at \url{https://github.com/mrjunjieli/wespeaker_u_cube}.}

\end{abstract}


\section{Introduction}

Speaker verification (SV) determines whether enrollment and test utterances belong to the same speaker and supports applications such as biometric authentication, surveillance, and personalized services \cite{wang2024overview}. Modern systems extract fixed-dimensional speaker embeddings with a neural encoder and process them using a back-end. The back-end is more than a single similarity function: scoring measures the compatibility of two embeddings, score normalization compensates for trial-dependent score distributions, and calibration maps the resulting scores to a scale on which a common decision threshold can be applied.

Cosine similarity and probabilistic linear discriminant analysis (PLDA) are two widely used scoring back-ends \cite{ioffe2006probabilistic,prince2007probabilistic,wang2022scoring}. PLDA models within- and between-speaker variability using development data, whereas cosine scoring directly compares embedding directions without training an additional back-end model. With embeddings trained using large-margin classification objectives, the field has increasingly shifted toward cosine scoring because the training objective already encourages angular speaker separation \cite{wang2022scoring}. This does not make cosine universally superior to PLDA; rather, it provides a simple and well-matched baseline for the large-margin embeddings considered in this work.

Raw scores remain sensitive to speaker and acoustic conditions even when the embedding extractor is strong. Score normalization estimates a reference distribution from impostor cohorts to reduce this variability. Z-norm normalizes with respect to the enrollment model, whereas T-norm operates on the test side \cite{auckenthaler2000score}. ZT-norm applies these transformations sequentially \cite{auckenthaler2000score,shum2010unsupervised}. Shum et al. introduced symmetric normalization (S-norm), which symmetrically combines enrollment- and test-side normalization \cite{shum2010unsupervised}. Adaptive S-norm (AS-Norm) further selects the most relevant cohort scores for each trial and has become a common choice in neural speaker verification systems \cite{karam2011towards,cumani11_interspeech}. We therefore use AS-Norm as a representative trial-adaptive normalization method, rather than assuming that it is optimal under every evaluation condition.

Normalization and calibration address related but distinct problems. Cohort normalization can improve score comparability and discrimination, but it does not guarantee well-calibrated likelihood ratios and may leave trial-dependent calibration errors \cite{cumani2022impostor}. Conventional logistic-regression calibration applies a global affine transformation, while condition- or quality-aware variants introduce side information to adapt the decision mapping. In particular, QMF-based calibration uses duration, embedding magnitude, and impostor statistics as auxiliary quality indicators when transforming the input score \cite{thienpondt2021idlab,cumani2022impostor}. We select QMF because it complements AS-Norm and provides an explicit interface through which additional reliability information can enter the final back-end stage.

However, both conventional AS-Norm and QMF treat speaker embeddings and cohort statistics as deterministic, although real-world utterances exhibit different levels of reliability due to duration, noise, channel, and recording conditions. Existing uncertainty-aware speaker models mainly improve front-end representation learning or the initial similarity score \cite{lee2021xi,wang2023incorporating,wang2024cosine,li2025xi+,li2026u3xipushingboundariesspeaker,chen2026uncertaintyfactorization,jin2026uncertaintyaware}. The resulting uncertainty is generally not propagated into the subsequent cohort normalization and calibration stages.

In this paper, we address this gap by presenting a unified uncertainty-aware back-end that takes paired embeddings and covariance matrices as inputs. The framework comprises uncertainty-aware cosine scoring, uncertainty-aware AS-Norm (UAS-Norm), and uncertainty-aware QMF calibration (UQMF). UAS-Norm adjusts cohort statistics and normalized-score combination according to embedding uncertainty, while UQMF incorporates covariance-adjusted quality measures into calibration. This scoring--normalization--calibration pipeline retains trial-dependent reliability throughout back-end processing without redesigning the speaker encoder.

\section{Unified Uncertainty-Aware Back-End Processing}
\label{sec:backend}
This section presents the unified uncertainty-aware back-end, covering cosine scoring, AS-Norm, and QMF calibration in sequence. We first summarize the embedding and covariance supplied to these stages.

We use the $\mathcal{U}^3$-xi framework \cite{li2026u3xipushingboundariesspeaker} to obtain an embedding and its uncertainty. Given frame-level representations $\{\mathbf{z}_t\}_{t=1}^{T}$, it predicts diagonal precision matrices $\mathbf{L}_t$ and produces the posterior
\begin{equation}
p(\mathbf{h}\mid\mathbf{z}_{1:T})
=\mathcal{N}(\mathbf{h}\mid\boldsymbol{\phi},\mathbf{L}^{-1}),
\qquad
\mathbf{L}^{-1}=\left(\sum_{t=1}^{T}\mathbf{L}_t+\mathbf{L}_p\right)^{-1},
\label{eq:posterior_uncertainty}
\end{equation}
where $\mathbf{L}_p$ is the prior precision, $\boldsymbol{\phi}$ is the posterior mean, and $\mathbf{L}^{-1}$ is the posterior covariance. After batch normalization (BN) and the final fully connected (FC) layer, the mean becomes the speaker embedding $\boldsymbol{\phi}^{\mathrm{s}}$, and the covariance becomes \cite{wang2023incorporating,li2025xi+,li2026u3xipushingboundariesspeaker}
\begin{equation}
\boldsymbol{\Sigma}^{\mathrm{s}}=
\mathbf{A}_{\mathrm{fc}}
\frac{\mathbf{L}^{-1}\odot\boldsymbol{\gamma}_{\mathrm{bn}}^2}
{\boldsymbol{\sigma}_{\mathrm{bn}}+\epsilon\mathbf{I}}
\mathbf{A}_{\mathrm{fc}}^{\top},
\label{eq:embedding_covariance}
\end{equation}
where $\mathbf{A}_{\mathrm{fc}}$ is the FC weight matrix; $\boldsymbol{\gamma}_{\mathrm{bn}}$ and $\boldsymbol{\sigma}_{\mathrm{bn}}$ are the BN scale and variance; $\odot$ denotes element-wise multiplication; and $\epsilon$ ensures numerical stability. We retain the diagonal of $\boldsymbol{\Sigma}^{\mathrm{s}}$ as direction-dependent embedding variances, with larger values indicating greater uncertainty. Thus, each utterance enters the back-end as the mean--variance pair $(\boldsymbol{\phi}^{\mathrm{s}},\boldsymbol{\Sigma}^{\mathrm{s}})$, and the operations below are evaluated element-wise.

\subsection{Cosine Scoring with Uncertainty}
\label{sec:ucos}
Given enrollment and test embeddings $\boldsymbol{\phi}_e^\text{s}$ and $\boldsymbol{\phi}_t^\text{s}$, conventional cosine scoring computes
\begin{equation}
s_{\text{cos-}o}(\boldsymbol{\phi}_e^\text{s},\boldsymbol{\phi}_t^\text{s})
= \frac{\langle \boldsymbol{\phi}_e^\text{s},\boldsymbol{\phi}_t^\text{s} \rangle}
{\lVert \boldsymbol{\phi}_e^\text{s} \rVert \, \lVert \boldsymbol{\phi}_t^\text{s} \rVert},
\label{eq:cos}
\end{equation}
where $\langle\cdot,\cdot\rangle$ and $\lVert\cdot\rVert$ denote the inner product and $\ell_2$ norm, respectively. To incorporate uncertainty, the Euclidean norms are replaced with covariance-adjusted effective norms while the inner product remains unchanged \cite{wang2024cosine,li2026u3xipushingboundariesspeaker}:
\begin{equation}
s_{\text{cos-}u}
= \frac{\langle \boldsymbol{\phi}_e^\text{s},\boldsymbol{\phi}_t^\text{s} \rangle}
{\sqrt{(\boldsymbol{\phi}_e^\text{s})^\top (\mathbf{I}+\boldsymbol{\Sigma}_e^\text{s})^{-1}\boldsymbol{\phi}_e^\text{s}}
 \sqrt{(\boldsymbol{\phi}_t^\text{s})^\top (\mathbf{I}+\boldsymbol{\Sigma}_t^\text{s})^{-1}\boldsymbol{\phi}_t^\text{s}}},
\label{equ:ucos}
\end{equation}
where $\boldsymbol{\Sigma}_e^\text{s}$ and $\boldsymbol{\Sigma}_t^\text{s}$ are obtained from (\ref{eq:embedding_covariance}). Equivalently, $s_{\text{cos-}u}=g_e g_t s_{\text{cos-}o}$, where
\begin{equation*}
g_a = \sqrt{
\frac{(\boldsymbol{\phi}_a^\text{s})^\top\boldsymbol{\phi}_a^\text{s}}
{(\boldsymbol{\phi}_a^\text{s})^\top
(\mathbf{I}+\boldsymbol{\Sigma}_a^\text{s})^{-1}
\boldsymbol{\phi}_a^\text{s}}},
\quad a\in\{e,t\}, \quad g_a\geq1.
\end{equation*}
The factors $g_e$ and $g_t$ provide trial-dependent score rescaling and are reused in UAS-Norm.

Because $\boldsymbol{\Sigma}_a^\text{s}$ is positive semidefinite, $(\mathbf{I}+\boldsymbol{\Sigma}_a^\text{s})^{-1}$ cannot increase the quadratic norm, which gives $g_a\geq1$. Uncertainty aligned with the embedding direction therefore produces a stronger scale adjustment than uncertainty concentrated in orthogonal directions. When $\boldsymbol{\Sigma}_a^\text{s}=\mathbf{0}$, $g_a=1$ and the formulation reduces to conventional cosine scoring. The dot-product similarity remains unchanged, while the covariance affects the score only through the normalization terms.

\subsection{UAS-Norm: Uncertainty-Aware AS-Norm}
\label{sec:uas-norm}
AS-Norm \cite{karam2011towards,cumani11_interspeech} reduces condition-dependent score variability by normalizing a trial against adaptively selected impostor cohorts. For side $a\in\{e,t\}$, let $\mathcal{C}_a^{(o)}=\{\mathbf{c}_{a,n}^{(o)}\}_{n=1}^{N}$ contain the top-$N$ cohort embeddings selected using $s_{\text{cos-}o}$. The conventional cohort mean and standard deviation are
\begin{equation}
\mu_o(\mathcal{C}_a^{(o)}) = \frac{1}{N}
\sum_{n=1}^{N}
s_{\text{cos-}o}(\boldsymbol{\phi}_a^\text{s},
\mathbf{c}_{a,n}^{(o)}),
\label{eq:as_mean}
\end{equation}
\begin{equation}
\sigma_o(\mathcal{C}_a^{(o)}) =
\sqrt{\frac{1}{N}\sum_{n=1}^{N}
\left(s_{\text{cos-}o}(\boldsymbol{\phi}_a^\text{s},
\mathbf{c}_{a,n}^{(o)})-\mu_o(\mathcal{C}_a^{(o)})\right)^2}.
\label{eq:as_std}
\end{equation}
The conventional AS-Norm score is
\begin{equation}
s_{\text{AS-}o} =
\frac{1}{2}
\left(
\frac{s_{\text{cos-}o} - \mu_o(\mathcal{C}_t^{(o)})}
{\sigma_o(\mathcal{C}_t^{(o)})}
+
\frac{s_{\text{cos-}o} - \mu_o(\mathcal{C}_e^{(o)})}
{\sigma_o(\mathcal{C}_e^{(o)})}
\right).
\label{eq:asnorm}
\end{equation}
Although this procedure adapts to the local score distribution, it treats every selected cohort embedding as equally reliable. We propose \textbf{uncertainty-aware AS-Norm (UAS-Norm)}, which introduces covariance into the cohort mean, cohort standard deviation, and final symmetric score combination.

The three components address complementary effects: unreliable cohort samples on the distribution center, their influence on its spread, and enrollment/test uncertainty in the final score. Their cumulative effects are evaluated in Table~\ref{tab:result}.

\subsubsection{Uncertainty-Aware Cohort Mean}
For the uncertainty-aware pipeline, let $\mathcal{C}_a^{(u)}=\{(\mathbf{c}_{a,n}^{(u)},\boldsymbol{\Sigma}_{a,n}^{(u)})\}_{n=1}^{N}$ contain the top-$N$ cohort embedding--covariance pairs selected using $s_{\text{cos-}u}$. Here, $a$ identifies the enrollment or test side and $n$ indexes the selected cohort samples. The original mean in (\ref{eq:as_mean}) assigns equal weight to every cohort embedding. We instead define the uncertainty-based reliability weight
\begin{equation}
w_{a,n} =
\frac{1}{
(\mathbf{c}_{a,n}^{(u)})^\top
\boldsymbol{\Sigma}_{a,n}^{(u)}
\mathbf{c}_{a,n}^{(u)}+\epsilon},
\qquad a\in\{e,t\},
\label{eq:cohort_weight}
\end{equation}
where $\boldsymbol{\Sigma}_{a,n}^{(u)}$ is the covariance of $\mathbf{c}_{a,n}^{(u)}$. A larger projected uncertainty produces a smaller weight. The uncertainty-aware mean is
\begin{equation}
\mu_u(\mathcal{C}_a^{(u)}) =
\frac{
\sum_{n=1}^{N}w_{a,n}
s_{\text{cos-}u}(\boldsymbol{\phi}_a^\text{s},
\mathbf{c}_{a,n}^{(u)})}
{\sum_{n=1}^{N}w_{a,n}}.
\quad \text{\textcircled{1}}
\label{eq:u_mean}
\end{equation}
The quadratic form measures uncertainty along the cohort direction, while normalization by $\sum_n w_{a,n}$ preserves the scale of the mean. Both the trial score and cohort reference scores use $s_{\text{cos-}u}$, while $w_{a,n}$ further controls the contribution of each cohort sample according to its uncertainty. Consequently, uncertainty-aware cosine scoring is used consistently throughout UAS-Norm and UQMF.

\subsubsection{Uncertainty-Aware Cohort Standard Deviation}
The uncertainty-aware cohort standard deviation is computed using the same reliability weights
\begin{equation}
\sigma_u(\mathcal{C}_a^{(u)}) =
\sqrt{
\frac{
\sum_{n=1}^{N}w_{a,n}
\Big(s_{\text{cos-}u}(\boldsymbol{\phi}_a^\text{s},
\mathbf{c}_{a,n}^{(u)})
-\mu_u(\mathcal{C}_a^{(u)})\Big)^2}
{\sum_{n=1}^{N}w_{a,n}}
}.
\quad \text{\textcircled{2}}
\end{equation}
Setting $a=e$ or $a=t$ gives the enrollment- and test-side statistics, respectively. Thus, uncertain cohort embeddings contribute less to both local distribution estimates.

Using the same weights keeps the center and spread consistent with the same uncertainty-weighted cohort distribution.

\subsubsection{Uncertainty-Aware Score-Scaled Combination}
\label{sec:uas-combination}
For the final combination, we reuse the scale factors $g_e$ and $g_t$ from Section~\ref{sec:ucos}. Their behavior complements the reliability weights: $w_{a,n}$ suppresses uncertain cohort samples when estimating reference statistics, whereas $g_a$ directly rescales the normalized trial score. The final UAS-Norm score is
\begin{equation}
s_{\text{AS-}u} =
g_t \frac{s_{\text{cos-}u} - \mu_u(\mathcal{C}_t^{(u)})}{\sigma_u(\mathcal{C}_t^{(u)})}
+
g_e \frac{s_{\text{cos-}u} - \mu_u(\mathcal{C}_e^{(u)})}{\sigma_u(\mathcal{C}_e^{(u)})}.
\quad \text{\textcircled{3}}
\label{eq:uasnorm}
\end{equation}
The factor $1/2$ in conventional AS-Norm is omitted because a positive global scale does not affect score ranking and can be absorbed by calibration. This is uncertainty-conditioned scale compensation rather than confidence averaging: the cohort weights suppress unreliable references, whereas $g_e$ and $g_t$ rescale the trial terms according to the uncertainty-aware cosine geometry.

\subsection{UQMF: Uncertainty-Aware QMF Calibration}
\label{sec:qmf_background}
QMF-based calibration augments the input score with trial-level quality measures \cite{thienpondt2021idlab}. Using logistic regression, the calibrated score is
\begin{equation}
s_{\text{QMF}}=w_s s_{\mathrm{in}}+\mathbf{w}_q^\top\mathbf{q}+b,
\label{eq:qmf}
\end{equation}
where $s_{\mathrm{in}}$ is the score to be calibrated, $\mathbf{q}$ is a quality vector, and $w_s$, $\mathbf{w}_q$, and $b$ are learned calibration parameters. In our WeSpeaker-based implementation \cite{wang2023wespeaker,wang2024advancing}, $\mathbf{q}$ contains duration-, embedding-, and impostor-based measures. For side $a\in\{e,t\}$, the conventional embedding measure is
\begin{equation}
q_{m_a\text{-}o}=
\sqrt{(\boldsymbol{\phi}_a^\text{s})^\top\boldsymbol{\phi}_a^\text{s}},
\end{equation}
while $\mu_o(\mathcal{C}_e^{(o)})$ and $\mu_o(\mathcal{C}_t^{(o)})$ serve as the impostor measures. Let $\mathbf{q}_o$ collect these measures together with the two duration measures. The conventional QMF-calibrated score is
\begin{equation}
s_{\text{QMF-}o}
=w_{s,o} s_{\text{AS-}o}
+\mathbf{w}_{q,o}^\top\mathbf{q}_o+b_o.
\label{eq:qmf_o}
\end{equation}

Conventional QMF again treats all quality inputs as deterministic. In the proposed \textbf{uncertainty-aware QMF (UQMF)}, the duration measures remain unchanged, the impostor measures use $\mu_u(\mathcal{C}_e^{(u)})$ and $\mu_u(\mathcal{C}_t^{(u)})$ from UAS-Norm, and the embedding measures are replaced with covariance-adjusted norms
\begin{equation}
q_{m_a\text{-}u}
=
\sqrt{
(\boldsymbol{\phi}_a^\text{s})^\top
(\boldsymbol{\Sigma}_a^\text{s} + \mathbf{I})^{-1}
\boldsymbol{\phi}_a^\text{s}
},
\qquad a\in\{e,t\}.
\end{equation}
The identity term regularizes the covariance adjustment. As uncertainty increases, the covariance-adjusted norm generally decreases, allowing it to indicate lower embedding quality. Unlike the scale factors in Section~\ref{sec:uas-combination}, this quantity is supplied as a calibration feature and its contribution is learned from calibration trials.

Let $\mathbf{q}_u$ collect $q_{m_e\text{-}u}$, $q_{m_t\text{-}u}$, $\mu_u(\mathcal{C}_e^{(u)})$, and $\mu_u(\mathcal{C}_t^{(u)})$ together with the duration measures. Substituting $s_{\text{AS-}u}$ and $\mathbf{q}_u$ into (\ref{eq:qmf}) gives
\begin{equation}
s_{\text{QMF-}u}
=
w_{s,u} s_{\text{AS-}u}
+
\mathbf{w}_{q,u}^\top\mathbf{q}_u+b_u.
\end{equation}
The conventional parameters $(w_{s,o},\mathbf{w}_{q,o},b_o)$ and uncertainty-aware parameters $(w_{s,u},\mathbf{w}_{q,u},b_u)$ are trained separately using $\mathbf{q}_o$ and $\mathbf{q}_u$, respectively. No additional uncertainty rule is imposed on the calibrated score: the signs and magnitudes of covariance-derived features are learned from calibration trials. At inference time, UQMF only requires the covariances and cohort statistics computed by the preceding stages and therefore does not require an additional speaker encoder.

\subsection{Implementation Considerations}
The proposed back-end does not require encoder retraining once covariances are available. Covariance-adjusted norms, scale factors, and projected cohort uncertainties are utterance-level quantities that can be cached across trials. UAS-Norm has no learnable parameters, while UQMF retains the logistic-regression structure of conventional QMF and changes only its inputs. The methods can therefore be attached to an existing normalization and calibration pipeline without changing the enrollment protocol.

\section{Experiments and Results}
\begin{table*}[t]
\setlength{\tabcolsep}{6pt}
\renewcommand{\arraystretch}{1.22}
\centering

\caption{Evaluation results using different back-end scoring methods and uncertainty configurations. The conventional pipeline follows $s_{\text{cos-}o}\!\rightarrow\!s_{\text{AS-}o}\!\rightarrow\!s_{\text{QMF-}o}$, while the uncertainty-aware pipeline follows $s_{\text{cos-}u}\!\rightarrow\!s_{\text{AS-}u}\!\rightarrow\!s_{\text{QMF-}u}$. The best value in each metric column is shown in \textbf{bold}, and the second-best value is \underline{underlined}. The ranking is performed separately for the ECAPA and ResNet groups. The UAS-Norm $s_{\text{AS-}u}$ incorporates all three uncertainty components (\text{\textcircled{1}+\textcircled{2}+\textcircled{3}}). ``+xi'' denotes the $\mathcal{U}^3$-xi uncertainty module, and $\dagger$ denotes results obtained directly from pretrained WeSpeaker models. RI is the average relative reduction over the corresponding architecture-specific benchmark across EER and minDCF on Vox1-O, Vox1-E, and Vox1-H.}

\resizebox{\textwidth}{!}{%
\begin{tabular}{c | c | c | c c c | c c | c c | c c | c}
\toprule
Row 
& Model
& \# Param. 
& Cosine 
& \multirow{2}{*}{AS-Norm} 
& \multirow{2}{*}{QMF} 
& \multicolumn{2}{c|}{Vox1-O} 
& \multicolumn{2}{c|}{Vox1-E} 
& \multicolumn{2}{c|}{Vox1-H} 
& \multirow{2}{*}{RI (\%)} \\
\cline{7-12}
& & & score & & 
& EER (\%) & minDCF
& EER (\%) & minDCF
& EER (\%) & minDCF
& \\
\hline

1 
& ECAPA$\dagger$ 
& 6.19 M 
& $s_{\text{cos-}o}$ 
& & 
& 1.069 & 0.122 
& 1.209 & 0.136 
& 2.310 & 0.226 
& Benchmark 
\\ 
\hline 
2
& \multirow{8}{*}{ECAPA+xi}
& \multirow{8}{*}{6.69 M}
& $s_{\text{cos-}o}$
& &
&
0.936 & 0.100
& 1.054 & 0.123
& 1.978 & 0.195
& 13.49
\\

3
& &
& $s_{\text{cos-}o}$
& $s_{\text{AS-}o}$
& &
0.840 & 0.111
& 0.979 & 0.115
& 1.809 & 0.173
& 18.34
\\

4
& &
& $s_{\text{cos-}o}$
& $s_{\text{AS-}o}$
& $s_{\text{QMF-}o}$
&
0.766 & 0.106
& 0.932 & 0.108
& 1.693 & \underline{0.167}
& 22.96
\\

\cline{4-13}

5
& &
& $s_{\text{cos-}u}$
& &
&
0.840 & \textbf{0.086}
& 0.965 & 0.110
& 1.833 & 0.189
& 21.21
\\

6
& &
& $s_{\text{cos-}u}$
& $s_{\text{AS-}u}(\text{\textcircled{1}})$
& &
\underline{0.761} & 0.101
& 0.913 & 0.104
& 1.677 & \underline{0.167}
& 24.59
\\

7
& &
& $s_{\text{cos-}u}$
& $s_{\text{AS-}u}(\text{\textcircled{1}+\textcircled{2}})$
& &
\underline{0.761} & 0.100
& 0.912 & 0.104
& 1.675 & \textbf{0.166}
& 24.83
\\

8
& &
& $s_{\text{cos-}u}$
& $s_{\text{AS-}u}(\text{\textcircled{1}+\textcircled{2}+\textcircled{3}})$
& &
\textbf{0.750} & 0.098
& \underline{0.906} & \underline{0.100}
& \underline{1.663} & \underline{0.167}
& 25.86
\\

9
& &
& $s_{\text{cos-}u}$
& $s_{\text{AS-}u}$
& $s_{\text{QMF-}u}$
&
\textbf{0.750} & \underline{0.095}
& \textbf{0.892} & \textbf{0.090}
& \textbf{1.629} & \textbf{0.166}
& 28.01
\\ \hline \hline

10
& ResNet$\dagger$ 
& 6.63 M 
& $s_{\text{cos-}o}$ 
& & 
& 0.867 & 0.091 
& 1.049 & 0.121 
& 1.960 & 0.192 
& Benchmark 
\\ 
\hline

11
& \multirow{6}{*}{ResNet+xi}
& \multirow{6}{*}{7.92 M}
& $s_{\text{cos-}o}$
& &
&
0.904 & 0.070
& 0.933 & 0.098
& 1.658 & 0.165
& 13.06
\\

12
& &
& $s_{\text{cos-}o}$
& $s_{\text{AS-}o}$
& &
0.888 & 0.079
& 0.922 & 0.096
& 1.618 & 0.163
& 12.68
\\

13
& &
& $s_{\text{cos-}o}$
& $s_{\text{AS-}o}$
& $s_{\text{QMF-}o}$
&
0.782 & 0.065
& 0.842 & \underline{0.090}
& 1.489 & \underline{0.151}
& 21.52
\\

\cline{4-13}

14
& &
& $s_{\text{cos-}u}$
& &
&
0.813 & 0.075
& 0.847 & 0.091
& 1.532 & 0.167
& 17.12
\\

15
& &
& $s_{\text{cos-}u}$
& $s_{\text{AS-}u}$
& &
\underline{0.771} & \textbf{0.052}
& \underline{0.805} & \textbf{0.086}
& \underline{1.400} & \textbf{0.145}
& 26.53
\\

16
& &
& $s_{\text{cos-}u}$
& $s_{\text{AS-}u}$
& $s_{\text{QMF-}u}$
&
\textbf{0.745} & \underline{0.053}
& \textbf{0.793} & \textbf{0.086}
& \textbf{1.386} & \textbf{0.145}
& 27.15
\\

\bottomrule 
\end{tabular}
}

\label{tab:result}
\end{table*}








\subsection{Experimental settings}
The models are trained on VoxCeleb2 \cite{chung2018voxceleb2} and evaluated on the cleaned VoxCeleb1-O, VoxCeleb1-E, and VoxCeleb1-H trial lists \cite{nagrani2020voxceleb}. We evaluate ECAPA-TDNN \cite{desplanques20_interspeech} and ResNet backbones implemented in WeSpeaker. Data augmentation is applied using noise samples from the MUSAN corpus \cite{snyder2015musan} and room impulse responses from the RIR database \cite{ko2017study}.

We follow the VoxCeleb2 training pipeline provided by WeSpeaker \cite{wang2023wespeaker}. Baseline encoders use AAM-Softmax, whereas encoders with the uncertainty module use the inter- and intra-speaker-aware UAAM-Softmax strategy in \cite{li2026robustuncertainty}. This objective introduces an uncertainty-dependent scale to jointly optimize speaker discrimination and uncertainty estimation. Within each ``+xi'' group in Table~\ref{tab:result}, the same encoder is evaluated with different back-end configurations.

Models trained in this work use 150 epochs and 2-second audio segments. The base scale $s$ is 32; the angular margin increases gradually from 0 to 0.2 between epochs 20 and 40 and then remains constant. We average the parameters of the last 10 checkpoints. For AS-Norm, speaker centroids from VoxCeleb2-dev form the cohort pool, and the top 100 scores are selected independently for each side. Cohort selection and statistics use $s_{\text{cos-}o}$ in the conventional pipeline and $s_{\text{cos-}u}$ in the uncertainty-aware pipeline. QMF training trials are generated from VoxCeleb2-dev. Further model and training details are provided in \cite{li2026u3xipushingboundariesspeaker,li2026robustuncertainty}. We evaluate downstream SV performance using equal error rate (EER, \%) and normalized minimum detection cost (minDCF), with $P_{\mathrm{target}}=0.01$ and $C_{\mathrm{FA}}=C_{\mathrm{Miss}}=1$.

\subsection{Results and Analysis}
Table~\ref{tab:result} evaluates the benefits of uncertainty-aware learning and back-end processing. Compared with the pretrained ECAPA-TDNN and ResNet benchmarks (Rows~1 and~10), the uncertainty-aware encoders evaluated with the same conventional cosine scoring (Rows~2 and~11) achieve RIs of 13.49\% and 13.06\%, respectively. All ECAPA-TDNN metrics and five of the six ResNet metrics improve. These results demonstrate the effectiveness of uncertainty-aware learning in improving speaker embeddings for downstream SV, even before covariance information is explicitly used in the back-end.

Comparing Rows~2 and~5 and Rows~11 and~14 shows the effect of uncertainty-aware cosine scoring. It improves all ECAPA-TDNN metrics and all ResNet EERs, although the ResNet minDCF degrades slightly on Vox1-O and Vox1-H. The ResNet RI still increases from 13.06\% to 17.12\%.

Joint uncertainty-aware scoring and normalization improve all six metrics for both architectures: Row~8 raises the ECAPA-TDNN RI from 18.34\% to 25.86\% relative to Row~3, while Row~15 raises the ResNet RI from 12.68\% to 26.53\% relative to Row~12. Rows~6--8 further examine the cumulative UAS-Norm configurations: \textcircled{1}, \textcircled{1}+\textcircled{2}, and \textcircled{1}+\textcircled{2}+\textcircled{3}. Their RIs increase from 24.59\% to 24.83\% and then to 25.86\%, supporting the effectiveness of the three-component design and the benefits of progressively combining its strategies. 

\begin{figure}[tbp]
    \centering
    \captionsetup{skip=4pt}
    \subfloat[AS-Norm: $s_{\text{AS-}o}$]{%
        \includegraphics[width=0.45\linewidth]{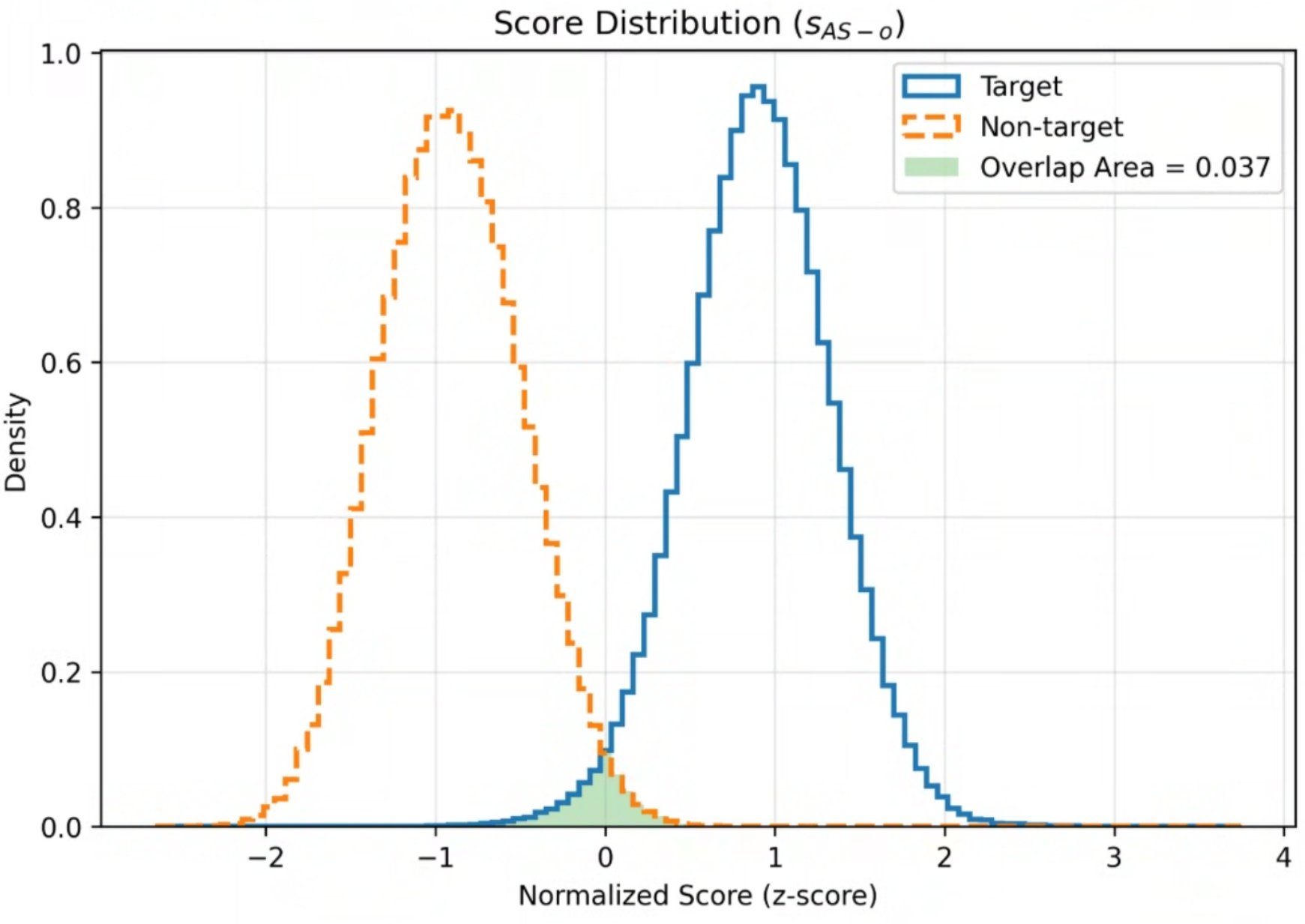}
    }
    \hfill
    \subfloat[UAS-Norm: $s_{\text{AS-}u}$]{%
        \includegraphics[width=0.45\linewidth]{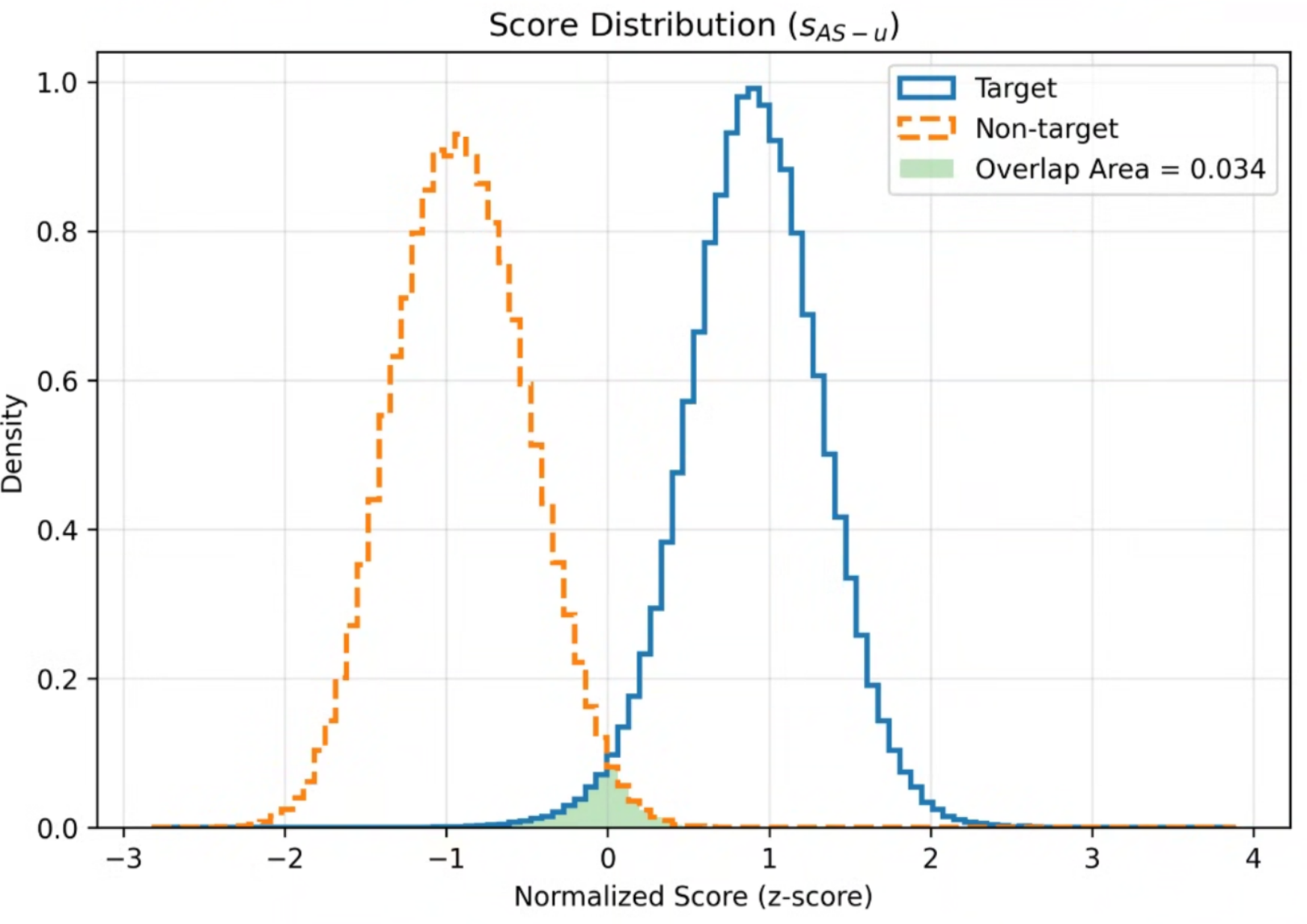}
    }
    \caption{Density distributions of the z-score-standardized conventional AS-Norm and uncertainty-aware AS-Norm scores on Vox1-H. Each class density integrates to one, and the shaded region denotes their overlap.}
    \label{fig:score}
\end{figure}

Figure~\ref{fig:score} compares the score distributions produced by conventional AS-Norm and the complete uncertainty-aware AS-Norm pipeline on Vox1-H. Because $s_{\text{AS-}o}$ and $s_{\text{AS-}u}$ have different raw scales, the scores of each method are transformed as $\widetilde{s}_j=(s_j-\mu_s)/\sigma_s$, where $\mu_s$ and $\sigma_s$ are computed from the pooled target and non-target scores of that method and then applied to both classes. The horizontal axis denotes the standardized score, while the vertical axis represents probability density. Each class density integrates to one, so the shaded overlap represents shared probability mass independently of the target--non-target trial imbalance. The complete uncertainty-aware pipeline decreases the overlap from 0.037 to 0.034, an 8.1\% relative reduction. This comparison reflects the aggregate effect of uncertainty-aware scoring, cohort statistics, and score combination rather than any single component.

The complete pipelines ending in UQMF (Rows~9 and~16) improve all six downstream SV metrics over their conventional counterparts ending in QMF (Rows~4 and~13). RI increases from 22.96\% to 28.01\% for ECAPA-TDNN and from 21.52\% to 27.15\% for ResNet. These are pipeline-level gains. The incremental comparisons (Rows~8--9 and~15--16) show that adding UQMF after UAS-Norm reduces five of the six EERs, with ECAPA-TDNN Vox1-O unchanged. ResNet Vox1-O minDCF increases slightly from 0.052 to 0.053. These results demonstrate the benefits of UQMF for downstream speaker verification.


\section{Conclusion}
We presented a unified uncertainty-aware back-end for speaker verification, comprising uncertainty-aware cosine scoring, UAS-Norm, and UQMF calibration. By reusing embedding covariance for score scaling, cohort weighting, and quality features, the framework preserves trial-dependent reliability throughout back-end processing. Experiments with ECAPA-TDNN and ResNet show consistent EER reductions and improved target--non-target separation, although occasional minDCF degradation indicates that the gains depend on uncertainty quality and the evaluation operating region.

\bibliographystyle{IEEEbib}
\bibliography{refs}

\end{document}